\documentclass[%
 reprint,
 amsmath,amssymb,
 aps,
]{revtex4-1}

\usepackage{graphicx}

\begin{document}

\preprint{APS/123-QED}

\title{The optimum conditions of laser field depletion in laser-electron beam collision}

\author{J. F. Ong}
\affiliation{%
 National Institute for Physics and Nuclear Engineering, ELI-NP, Str Reactorului, nr. 30, P.O.Box MG-6, Bucharest-Magurele 077125, Romania.}%
\affiliation{Research Center for Nuclear Physics, 10-1 Mihogaoka Ibaraki, 567-0047 Osaka, Japan}
\author{T. Moritaka}
\affiliation{Fundamental Physics Simulation Research Division, National Institute for Fusion Science,
322-6 Oroshi-cho, Toki City, GIFU Prefecture, 509-5292 Japan}
\author{H. Takabe}
\affiliation{Helmholtz-Zentrum Dresden-Rossendorf (HZDR), Bautzner Landstrasse 400, 01328 Dresden, Germany}

\date{\today}

\begin{abstract}
The laser field depletion in laser-electron beam collision is generally small. However, a properly chosen parameter of the laser and electron, the laser field depletion can be optimized. To access the laser energy evolution, simulations of laser-electron beam collision by Particle-in-Cell (PIC) simulation is performed. In this paper, the laser and electron parameters are chosen such that the ponderomotive force is compensated by the radiation reaction force in the head-on collision configuration. Then, the relativistic electron beam can quiver in the laser pulse for a longer time to increase the energies conversion. The optimum of laser field energy depletion is observed at $\gamma_0= a_0 \sim 400$ and limited beyond this point due to the impenetrability threshold. The total energy conversion from laser and electron beam into radiation emission is optimum at $\gamma_0= a_0 \sim 250$. This conversion efficiency can be up to several percent for an electron bunch with charges of a few nC. 
This efficient gamma-ray sources may provide some useful application in photonuclear experiments.
\end{abstract}

\pacs{Valid PACS appear here}
\maketitle


The experiment of laser-matter interaction at laser intensities of $10^{22}$ W/cm$^2$ has been achieved following the rapid progress in laser technology \cite{Nishiuchi:17}. Lasers with unprecedented intensities at $10^ {23}$ W/cm$^2$ are achievable in the coming years at 10-PW laser facilities under project \cite{ELI}. In line with the advancement in Laser Wakefield Acceleration (LWFA) \cite{PhysRevLett.43.267,Mangles,Geddes,Faure,Leemans} and the availability of linear accelerator (LINAC) at these facilities \cite{ELINP}, studies of the interaction of ultraintense laser pulse with a counter-propagating relativistic electron beam under extreme conditions are feasible. Under such conditions, the effects of radiation reaction (RR) become important and the production of intense gamma-ray can be obtained \cite{PhysRevLett.102.254802,PhysRevX.2.041004,PhysRevLett.112.015001,MyPoP}.    

The conditions for of laser-electron beam collision can be characterized by three dimensionless parameters. The first parameter is the normalized laser amplitude $a_0=|e||\bold{E}|/(mc\omega_0)$, where $|\bold{E}|$ and $\omega_0$ are the amplitude and angular frequency of the laser field, respectively while $m$ and $-e$ are the electron rest mass and charge, respectively. The speed of light is $c$. An electron becomes relativistic from rest in one laser cycle when $a_0 >> 1$. The second parameter is the normalized electron energy $\gamma=\mathcal{E}/mc^2$, where $\mathcal{E}$ is the total energy of an electron. An electron is relativistic when $\gamma >> 1$. The third is the gauge invariant parameter $\chi_e=\hbar\sqrt{-f_{L}\cdot f_{L}}/(m^2c^3)$, where $\hbar$ is the reduced Planck's constant and $f_{L}^{\mu}=\gamma(\mathbf{f}_{Le}\cdot \mathbf{u}/c, \mathbf{f}_{Le})$, $\mathbf{f}_{Le}$ is the Lorentz force. This parameter measures the ratio of the field strength acted on the electron to the Schwinger field $E_S=m^2c^3/(e\hbar)$. When $\chi_e \sim 1$, electron recoil becomes substantial and radiation emission have to be modified.

A non-negligible amount of the laser energy is absorbed in the laser-electron collision at $a_0,\gamma>>1$ \cite{MyPoP,MyPoP2,PhysRevLett.118.154803}. The depletion of the laser field is found to be significant at $a_0\sim 10^3$ for electron bunch with charges of 10 nC \cite{PhysRevLett.118.154803}. Although, with the availability of such ultraintense laser and highly charged bunches, the laser-electron collision suffers from the ponderomotive repulsion even if the electron is initially relativistic. This effect is even robust for an ultraintense laser focused to wavelength-scale waist radius and lead to the suppression of the RR. In this situation, only a small fraction of energies absorbed by electrons are being converted into radiation and thus signaling the effort to produce efficient and collimated gamma-ray source \cite{MyPoP2}. 
 
In this paper, we reports the characterization of the laser and electron parameters such that the ponderomotive force is balanced by the RR force in laser-electron collision. Unlike the case in laser-plasma interaction, electrons are initially relativistic. Electrons are either reflected by the laser pulse if $\gamma<<a_0$ or passing through if $\gamma>>a_0$ \cite{PhysRev.150.1060,Narozhny2000}, and leading to a shorter interaction time in the high-intensity region. When the designate characterizations are met, we observed that the electrons spend extra laser cycles quivering in the laser pulse as compared to the case without RR. The optimum of laser field depletion and conversion efficiency are observed via PIC simulations. At the optimum point, the effects of RR enhance the laser energy depletion by a few times which is subsequently converted into gamma-ray emission. We then propose a self-synchronized and all-optical setup to generate multiple gamma-ray sources by invoking the RR in a single laser pulse. This provides an impact to the photonuclear experiments which required a bright and collimated gamma-ray source with energies in the range between 10 - 30 MeV \cite{Ejiri,PhysRevSTAB.15.024701,RomRep}.
 
To circumvent the ponderomotive repulsion and increase the laser-electron interaction time, a few approaches were proposed. This involves the utilization of other forces to compensate the ponderomotive force. For the case of plasma, this proceeds through the so-called Radiation Reaction Trapping (RRT) \cite{PhysRevLett.112.145003}. At a certain intensity threshold, the RR force is strong enough to balance the ponderomotive force. The deviation of the electron motion in the transverse direction reduces and then trapped in the laser pulse. The intensity threshold for RRT is $a_{thr}\sim(w_0/r_e)^{1/3}$, where $r_e=e^2/mc^2$ is the classical electron radius and $w_0$ is the laser waist radius. When the laser is focused to the order of laser wavelength, the threshold is $a_{thr}\sim700$. In this regime, the electron spends a longer time oscillating inside the laser pulse and the laser energy conversion for radiation emission as well as electron-positron pair production is promising \cite{1367-2630-17-5-053039,Ncomm713686}. Another approach is to utilize the induced space-charged field by the laser pulse. At a particular plasma density, the space-charged field becomes large enough to balance the radiation pressure of the laser and hold the electrons in the laser pulse. This approach reduces the threshold power needed for quantum electrodynamic (QED) avalanche \cite{NSciRep715302}.

Unlike the RRT in the laser-plasma interaction, electron bunch is initially relativistic. Taking into account the effects of RR, the oscillating amplitude of a relativistic electron in the laser pulse becomes large due to the decrease of electron effective mass ($y\propto 1/(\gamma m)$) \cite{MyPoP}, which is in contrary to the aforementioned case. Moreover, with the absence of ions, the space-charged field cannot be induced. The electron bunch is either being repelled away by the ponderomotive force or passing through the laser pulse, leading to a short interaction time in the high-intensity region.

To characterized the conditions of force balance in laser-electron bunch collision, we derived the ratio of the magnitude of RR force to the ponderomotive force. The force balance is met if 
\begin{eqnarray}
	\frac{f_{RR}}{f_P}\sim \frac{2}{3}\pi\alpha\frac{\hbar\omega_0}{mc^2}\frac{w_0}{\lambda}\gamma^3 \sim 1, \label{frr/fp}
\end{eqnarray}
where $\alpha=1/137$ is the fine structure constant, $f_{RR}\sim 4\pi\alpha \hbar\omega_0\gamma^2a_0^2/(3\lambda)$ and $f_P\sim-mc^2\nabla a^2/(2\gamma)$ are the magnitude of the RR force and ponderomotive force, respectively \cite{MyPoP2}. Here, $\gamma=\sqrt{1+ <\mathbf{a}^2>}$, where $ \mathbf{a}=e\mathbf{A}/(mc^2)$ and $\mathbf{A}$ is the vector potential of the external field \cite{macchi2013superintense}. If the a Gaussian spatial profile is assumed for $ \mathbf{a}$, then the electron should have at least $\gamma\sim a_0 \sim 250$ to overcome the ponderomotive force from a laser pulse with $w_0=2\lambda$ and $\lambda=1~ \mu\text{m}$.  


Next, to optimize the energy transfer we analyzed the work done by laser in the context of classical electrodynamics. The work done per unit time per unit volume is expressed as $\bold{J}\cdot\bold{E}=-en_e\bold{u}\cdot\bold{E}$ where $n_e$ is the electron density and $\bold{u}$ is its velocity \cite{jackson1975classical}. Now, the laser is assumed to propagate in $+x$ direction and polarized in $y$ direction. By taking the momentum that couples to the laser polarization direction, the unit work done can be written as $\bold{J}\cdot\bold{E}\sim-en_e |\bold{p}_{\perp}| |\bold{E}_{\perp}|/(\gamma  m)$ where $|\bold{p}_{\perp}|$ is the momentum of the electron perpendicular to the laser propagation, i.e. parallel to the strongest electric field component. For an electron quivering in the electromagnetic field, its momentum can be approximated as $|\bold{p}_{\perp}|\sim a_0mc$ \cite{macchi2013superintense} provided the intensity variation is small over a wavelength. The unit work done is now   

\begin{eqnarray}
\bold{J}\cdot\bold{E} \propto -n_e\frac{a_0^2}{\gamma}mc^2\omega_0. \label{jdotE}
\end{eqnarray}

For $\gamma>>a_0$ and $a_0\sim 1$, it is certain that the energy conversion from the laser field to the electron is negligible. For $a_0>>\gamma$, laser energy is absorbed and only a small fraction being converted into radiation emission. However, under the condition $\gamma\sim a_0$, integrate the unit work over the volume and average over a laser frequency, the work done per particle per cycle is $W_{EM}\propto a_0 mc^2$. Then, the energy equation for a radiating electron can be written as 

\begin{eqnarray}
mc^2(\gamma - \gamma_0) = W_{EM} - W_{rad} \label{EnergyEq}
\end{eqnarray}
where $\gamma_0$ is the initial electron beam energy and $W_{rad}$ is the radiation energy. By letting $\gamma\sim a_0=\gamma_0$, the left hand side of Eq. (\ref{EnergyEq}) goes to zero and the work done by laser is almost converted into radiation energy. The energy conversion may be increased if the electrons trapped for a longer time in the laser pulse.

To validate the electron beam trapping and the laser field depletion, we performed a series of 3-dimensional PIC simulations of an electron beam interact with a counter-propagating laser pulse at $\gamma_0=a_0$. In these PIC simulations, we applied the Sokolov's model with QED modification \cite{SokolovMourou, SokolovJETP, SokolovModeling}. 
The size of the simulation box was $x\times y\times z = 30\lambda \times 10\lambda \times 10\lambda$ with $1200 \times 200 \times 200$ cells. We used $\text{1.25}\times 10^5$ macroparticles to represent an electron bunch containing $10^9$ electrons with the charge of $\sim$100 pC. We assumed an ideal electron beam with a Gaussian profile in phase space. The longitudinal and transverse electron beam sizes were 1 $\mu$m, respectively. A 5 $\%$ of momentum spread was used. A pulsed Gaussian laser beam beyond paraxial approximation up to 5$^{th}$ order correction was considered. The temporal profile was $g(\eta)=1/\cosh(\eta/\eta_0)$ \cite{parax}, where $\eta=\omega_0 t -kx$, $\eta_0=\omega_0t_L$, and $t_L$ is the laser pulse duration. In this work, the laser is polarized in $y$ direction and propagates along the $+x$ axis. The laser waist radius is $w_0=$2 $\mu$m with $\lambda=$ 1 $\mu$m, and a pulse duration of 10 fs. 

\begin{figure}[ht]
\centering
\includegraphics[width=\linewidth]{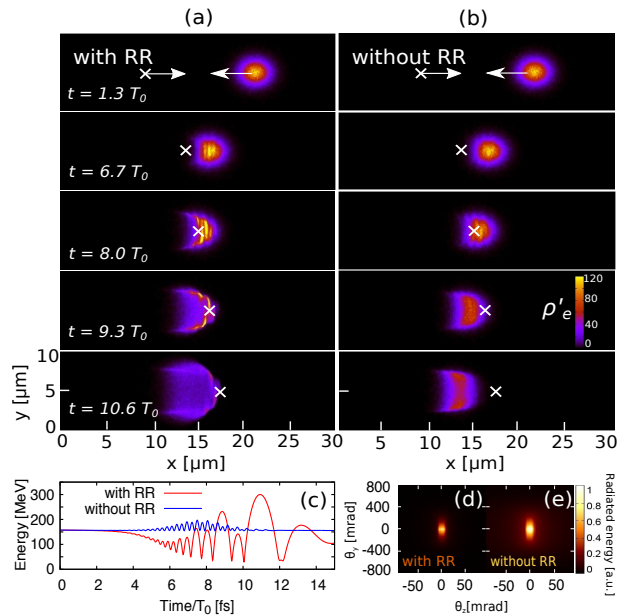}
\caption{The normalized electron density distribution $\rho'_e=4\pi^2\rho_e/en_{cr}$ of an electron beam with $\gamma_0=a_0=300$ for the case (a) with RR, and (b) without RR at $t=1.3T_0$, $t=6.7T_0$, $t=8.0T_0$, $t=9.3T_0$, and $t=10.6T_0$ ($T_0=3.3$ fs is the time for one laser cycle). The arrows represent the direction of the propagation of the laser and electron beam. The crosses represent the center of the laser pulse. (c) The energy evolution of a sampled electron in the beam. (d) $\&$ (e) The photon angular distributions with and without RR, respectively.}
\label{fig:density}
\end{figure}

In the absence of RR and $\gamma_0=a_0>>1$, the momentum components parallel and transverse to the laser beam axis are $p_x\sim |\bold{p}_{\perp}|\sim a_0mc$. The direction of the resultant momentum vectors are 45$^{\circ}$ with respect to the center of mass of the electron bunch in both y directions and eventually split apart. When the effects of RR is included, the RR force could balance the ponderomotive force to a certain extent and traps the electrons in the laser pulse. A comparison of the electron bunch density distribution between the case with and without RR is shown in Fig. \ref{fig:density} (a) $\&$ (b), where $\rho'_e=4\pi^2\rho_e/en_{cr}$ is the normalized electron density and $n_{cr}$ is the critical density.

As shown in Fig. \ref{fig:density} (a) at $t=8.0T_0$, a dense periodic electron structure starts to form at the center of the laser pulse with the period of the laser wavelength. The dense electron structure continues and moving along with the laser pulse until $t=9.3T_0$. Because of the small energy spread (5 $\%$ was assumed), some electrons at the edge of the bunch continue moving through the laser pulse. They form a boundary at the edge of the laser pulse instead of being expelled by the ponderomotive force. The electrons at the boundary of the laser field satisfied the penetrability condition, $\gamma_0 >a_0$. This boundary continues until $t=10.6T_0$ while a small part of the electrons still remain in the laser pulse.  

When the RR force is turned off, the dense electron structure does not form, however, an inconspicuous periodic structure was observed at $t=8.0T_0$ as shown in Fig. \ref{fig:density} (b). Then, the electron beam starts to be broken into two at an angle around 45$^{\circ}$ at $t=9.3T_0$ due to the absence of RR force to compensate the ponderomotive force. The electron beam then leaves the laser pulse after $t=10.6T_0$. The distinction of electron density between the case with and without RR only becomes obvious at $t=8.0T_0$ even if the pre-pulse of the laser start to interact with the electron beam before $t=6.7T_0$.

Figure \ref{fig:density} (c) shows the energy evolution of a sampled electron in the bunch. For the case without RR, there is no net change in the electron energy while a decrease in electron energy was observed for the case of RR. The phases correspond to energies absorption and emission from the laser to gamma-rays.  A time difference of $5T_0$ is observed between both cases. In other words, the sampled electron including the effects of RR spends an extra 5-cycle ($\sim17$ fs) in the laser pulse as compared to the case without RR which is about three times longer than the laser pulse duration. This will results in the increases of gamma-ray conversion efficiency as we will see later.

To assess the radiation angular distribution of an electron beam, we count and accumulate photons ($\chi\ge0.001$) at their emission points throughout the simulation. For a relativistic electron that moves in strong laser field, its radiation is emitted in the direction of electron momentum. When the RR force is absence, the electron bunch splits and moves in two directions. Therefore, one can observe the radiation emission is confined in a larger opening angle ($\theta_y\sim 300$ mrad $\sim 17^{\circ}$) as shown in Fig. \ref{fig:density} (e). Here, $\theta_{x,y}$ is the opening angle for the radiation emission in $x,y$ direction, respectively. As the direction of RR force is mostly in the direction of the Poynting vector \cite{0953-4075-31-3-002}, it confines the electrons bunch from splitting apart. As a result, the radiation is emitted in the forward direction within a smaller opening angle ($\theta_y\sim 200$ mrad $\sim 11^{\circ}$) as shown in Fig. \ref{fig:density} (d). Since the radiation are dominantly emitted in plane of laser polarization, so the opening angle in $z$ direction is $\theta_z\sim 12$ mrad ($\sim 0.7^{\circ}$) for both cases. The emission of gamma-ray is collimated within a narrow opening angle as compared to radiation generation by using plasma or solid target ($\sim 30^{\circ}$) \cite{PhysRevLett.112.145003,PhysRevLett.108.195001,1367-2630-17-5-053039}. 

\begin{figure}[ht]
\centering
\includegraphics[width=\linewidth]{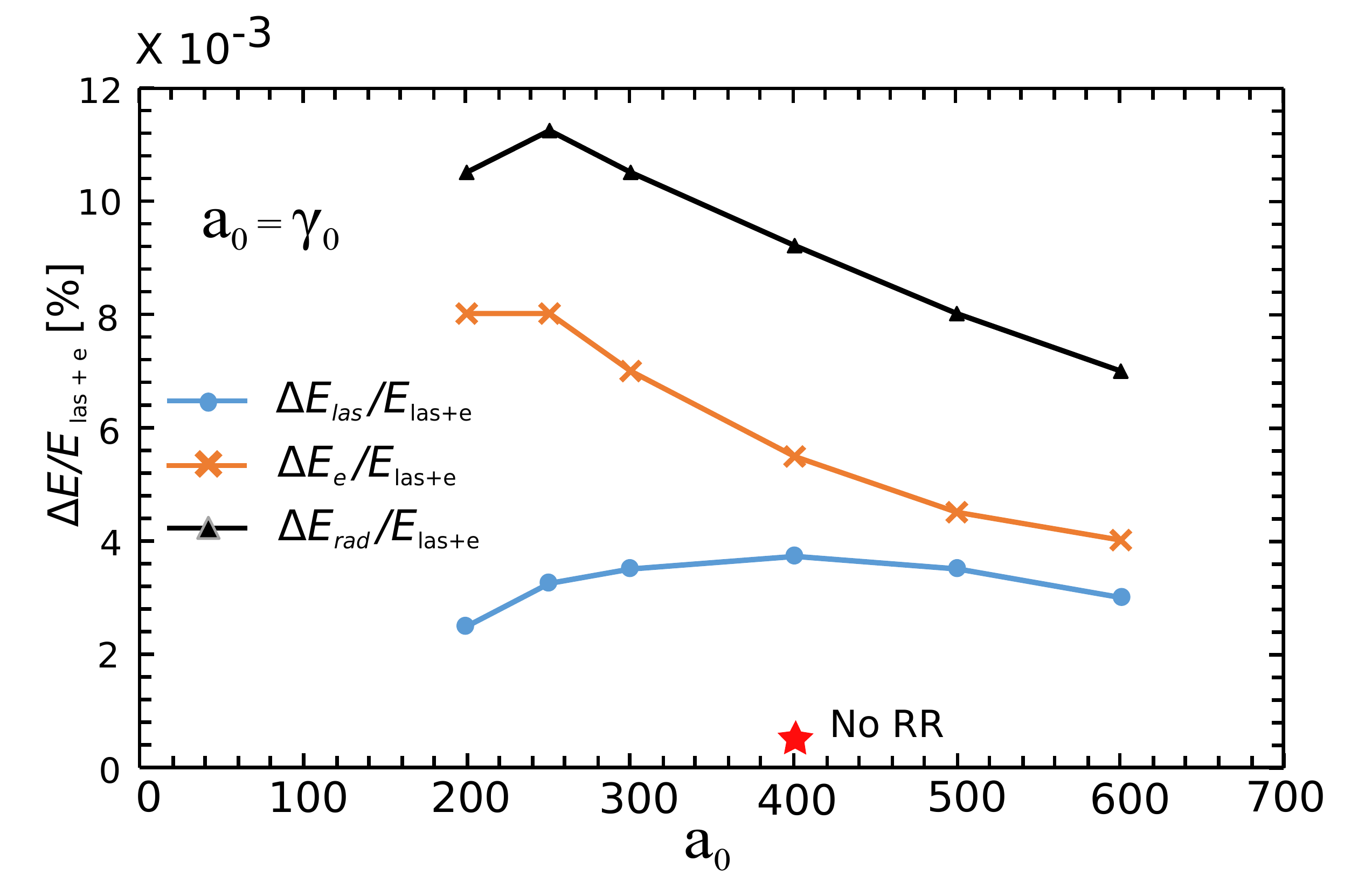}
\caption{The energy conversion efficiency of the electron, $\Delta E_e/E_{las+e}$, laser, $\Delta E_{las}/E_{las+e}$, and radiation energy, $\Delta E_{rad}/E_{las+e}$ for $\gamma_0=a_0=200, 250, 300, 400, 500,$ and $600$. The star correspond to $\Delta E_{las}/E_{las+e}$ for the case of no RR.}
\label{fig:a_equal_g}
\end{figure}

We carry on with the analysis on the energy conversion efficiencies of the electron $\Delta E_e/E_{las+e}$, laser $\Delta E_{las}/E_{las+e}$, and radiation $\Delta E_{rad}/E_{las+e}$ as shown in Fig. \ref{fig:a_equal_g}, where $E_{las+e}$ is the sum of laser and electron bunch energy.  Since $E_{las+e}$ is the initial value and constant, the lost of laser and electron beam energies go to the gamma-radiation. This means the that $\Delta E_e + \Delta E_{las}=\Delta E_{rad}$. We note that the values of $E_{las+e}$ are different for each $\gamma_0=a_0$. The laser energy depletion becomes large for increasing $\gamma_0=a_0$ but the conversion efficiencies might not be the case. For example, the laser energy depletion at $\gamma_0=a_0=400$ is 10 mJ while $\gamma_0=a_0=600$ is 19 mJ. However, the conversion efficiencies increase and reach an optimum point at $\gamma_0=a_0=400$. Beyond this point, it starts to decrease which is in contrary to Eq. \ref{jdotE}. This can be explained by the impenetrability of the electron into the laser pulse induced by the RR \cite{PhysRevA.90.053847}. Upon surpassing the impenetrability threshold, the electron bunch is not able to access the core of the laser pulse regardless of their initial kinetic energy. The laser energy depletion increases in small amount across a wide laser intensity range and resulting in the reduction of conversion efficiencies. The intensity threshold for the impenetrability is  $a_0\gtrsim a_{0*}(4\mu)^{1/3}$ where $a_{0*}=(3mc^2/2e^2\omega_0)^{1/3}\simeq 470$, and $\mu=\lambda/(\pi w_0)$ is the diffraction angle of Gaussian laser beam. In this work, the impenetrability threshold is $a_0\gtrsim $ 400 which is in agreement with the simulation result. This implies that the impenetrability threshold sets an upper limit to the laser field depletion efficiency. The laser field depletion efficiency becomes insignificant at very small and very high value of $a_0$. At the maximum point, we performed the same simulation without RR. We found that the RR enhanced the laser energy depletion by a factor of 6 as shown in Fig. \ref{fig:a_equal_g} by the star.

Apart from the laser field depletion, we also note the optimum of total radiation energy at $200\leq a_0\leq300$.  Since energy loss from electrons and laser are both converted into radiation emission, we can regard $\Delta E_{rad}/E_{las+e}$ as the energy conversion efficiency of the laser and electron into gamma-rays. This conversion efficiency scales with the power law of $a_0^{-7/9}$ between $a_0=300$ and $600$.

There are two considerations in selecting an optimum conversion efficiency$-$a large electron energy loss and a large laser field depletion. However, one cannot have both conditions at the same time. It is evident that the laser energy depletion is maximum at $a_0=400$ but not for the electron energy loss. For $a_0=200$, the electron energy loss is larger but the laser depletion is now small. Therefore, an ideal choice would be between $a_0=250~\&~300$.

The ratio of the RR force to the ponderomotive force for $200\leq a_0\leq400$ are listed in Table. \ref{Table1} with the corresponding radiation energy conversion efficiencies. Indeed, the maximum of the conversion efficiency happens to be at $f_{RR}/f_P\sim 1$ as the requirement of the compensation of the ponderomotive force by RR force.

\begin{table}
\centering
\caption{The ratio $\frac{f_{RR}}{f_P}$ at $200\leq a_0\leq400$ and the corresponding radiation energy conversion efficiencies.}
\begin{ruledtabular}
\begin{tabular}{lcr}
$\frac{f_{RR}}{f_P}$ & $\gamma_0=a_0$ & $\frac{\Delta E_{Rad}}{E_{las+e}} \times 10^{-3}[\%]$\\
\hline
0.478 & 200 & 10.5 \\
0.934 & 250 & 11.2 \\
1.615 & 300 & 10.5 \\
3.827 & 400 & 0.92 \\
\end{tabular}
\end{ruledtabular}
\label{Table1}
\end{table}%

The number of electron in plasma is obviously more than the electron bunch with charges of a few pC. Therefore one can obtain a larger conversion efficiency in plasma. In this work, the electron bunch charge is of the order of 100 pC and the energy conversion efficiency is small as evident in Fig. \ref{fig:a_equal_g} and Table. \ref{Table1}. On average, an electron in the bunch emits an energy of $1.2\times 10^{-11}$ J.  Recently, the production of an electron bunch with the charges of 100 nC at MeV-level energy is reported \cite{100nC}. They were produced by the interaction of TW-scale laser with a solid target at an oblique angle. If a second laser pulse is used for the head-on collision with such an electron bunch, the total emitted energy would be 12 J ($E_{las+e}=110$ J at $\gamma_0=a_0=250$). If we assume the laser-electron collision scale linearly with the electron number we can extrapolate to the case of 100 nC bunch and the conversion efficiency of 11$\%$ may be expected.

In the realistic situation, it is difficult to ensure a perfect synchronization of a tightly focused laser pulse with an electron bunch. To avoid this problem, an all-optical scheme was proposed. In this scheme, a driver laser pulse in LWFA is reflected by the plasma mirror and make a head-on collision with the accelerated electron bunch \cite{Phuoc}. The x-rays and gamma-rays generation has been achieved experimentally by using this scheme with the accelerated electron bunch with charges of a few pC \cite{Phuoc,Yu,Tsai}. However, proper choices of laser and plasma parameters could yield the electron bunch with charges of a few nC \cite{Martins}. Moreover, a plasma mirror may undergo surface modulation under the PW-class laser pulse. Then, the plasma mirror does not only reflect the driver laser pulse but also tightly focus to the spot size of a few laser wavelength with the increases in intensity up to ten times of the original driver pulse \cite{Tsai2}. Thus, we anticipate that a laser pulse with intensities $\sim10^{21}$ W/cm$^2$ would produce MeV-level electron bunch with charges of a few nC. Then, the back-reflected pulse would achieve an intensity boost up to $\sim10^{22}$ W/cm$^2$. At this time, the effects of RR would affect the laser-electron collision and a large amount of electron energy is being converted into gamma-rays. After the interaction, the electron bunch passes through the plasma mirror. If the plasma mirror is made up of high-Z material, the electron bunch will generate a second gamma-ray via Bremsstrahlung. Hence, a larger portion of the laser energy can be converted into gamma-rays.  

In summary, we have characterized the optimum conditions for laser field depletion such that the ponderomotive forces are balanced by the RR forces in transverse and longitudinal direction. As a result, the electrons spend a longer time quivering in the laser pulse as compared to the case without RR. We then observed that the laser field depletion was maximum at $\gamma_0=a_0=400$. Beyond this point, the field depletion was limited due to the impenetrability threshold induced by the RR. By combining the field depletion together with the electron beam energy lost, the optimum gamma-ray conversion efficiencies were found to be at $a_0=\gamma_0\in[250,300]$ with $f_{RR}/f_P\sim 1$. In this regime, the ponderomotive force was compensated by the RR force and resulting in the trapping of electrons in laser pulse. It is expected that an electron bunch with charges of 100 nC would convert up to 11 $\%$ of the combined laser and electron energy into radiation emission. The prospect of the generation of multiple gamma sources by using a single laser pulse was presented. In future work, we plan to study the generation of gamma-rays by using this setup. The efficient gamma source is of great interest for photonuclear isotope transmutation, nondestructive high-sensitivity detection of nuclear isotopes as well as the experiment of Production and Photoexcitation (PPEx) of isomers at Extreme Light Infrastructure (ELI-NP) \cite{Ejiri,PhysRevSTAB.15.024701,RomRep}.

\begin{acknowledgments}
This work is supported by Extreme Light Infrastructure $-$ Nuclear Physics (ELI-NP) $-$ Phase I, and also Phase II, a project co-financed by the Romanian Government and the European Union through the European Regional Development Fund. Simulations were performed at Large-Scale Computer System at Cybermedia Center Osaka University.
\end{acknowledgments}




\end{document}